\documentclass{osa-article}
\journal{osajournal}  

\newcolumntype{M}[1]{>{\centering\arraybackslash}m{#1}}
\usepackage{braket}
\usepackage{diagbox}
\usepackage{multirow}

\newcommand{\Tr}{\mathrm{Tr}}
\begin{document}

\title{Generation of quantum states with nonlinear squeezing by Kerr nonlinearity}

\author{\v{S}imon Br\"{a}uer\authormark{1}, Petr Marek\authormark{1*}}

\address{\authormark{1} Optics Department, Faculty of Science, Palack\'{y} University, 17. Listopadu 12,
 77146 Olomouc, Czech Republic}
\email{\authormark{*}marek@optics.upol.cz} 

\begin{abstract*}
Quantum states with nonlinear squeezing are a necessary resource for deterministic implementation of high-order quadrature phase gates that are, in turn, sufficient for advanced quantum information processing. We demonstrate that this class of states can be deterministically prepared by employing a single Kerr gate accompanied by suitable Gaussian processing. The required Kerr coupling depends on the energy of the initial system and can be made arbitrarily small. We also employ numerical simulations to analyze the effects of imperfections and to show to which extent can they be neglected.
\end{abstract*}
\section{Introduction}
Quantum information is an interdisciplinary field with the goal of utilizing quantum properties of various physical systems in order to answer the fundamental questions about nature and to drive new applications. The most promising areas of development are quantum communication \cite{Scarani2009Sep, Lo2014Aug, Azuma2015Dec, Ecker2021Jan}, expanding the tools of secure communication towards real-life application, quantum metrology \cite{Pryde2003Nov, Mitchell2004May, Aasi2013Aug}, practically enhancing the most powerful measurement tools of the modern times, and quantum computation \cite{Knill2001Jan, O'Brien2007Dec, Zhong2020Dec}, already approaching problems impossible for classical computers. Quantum information is not tied to any particular physical system. It aims at devising universal protocols, which then can be adapted by any capable platform. The protocols can be divided into two broad categories. The first one relies on discrete quantum systems represented by a specific number of two-level qubits. The second one considers continuous systems described by infinite dimensional Hilbert spaces. These can be the modes of optical \cite{Braunstein2005Jun, Weedbrook2012May, Andersen2015Sep} or microwave fields \cite{Ofek2016Aug, Hillmann2020Oct}, or the vibrational modes of mechanical oscillators \cite{Aspelmeyer2014Dec}. Alternatively they can be also systems of qubits of such a large number that their collective behavior is essentially continuous, such as in the case of collective magnetic spins \cite{Hosten2016Jan, Cuevas2017Sep}.
\begin{figure}[h!]
\centerline{\includegraphics[width=12cm]{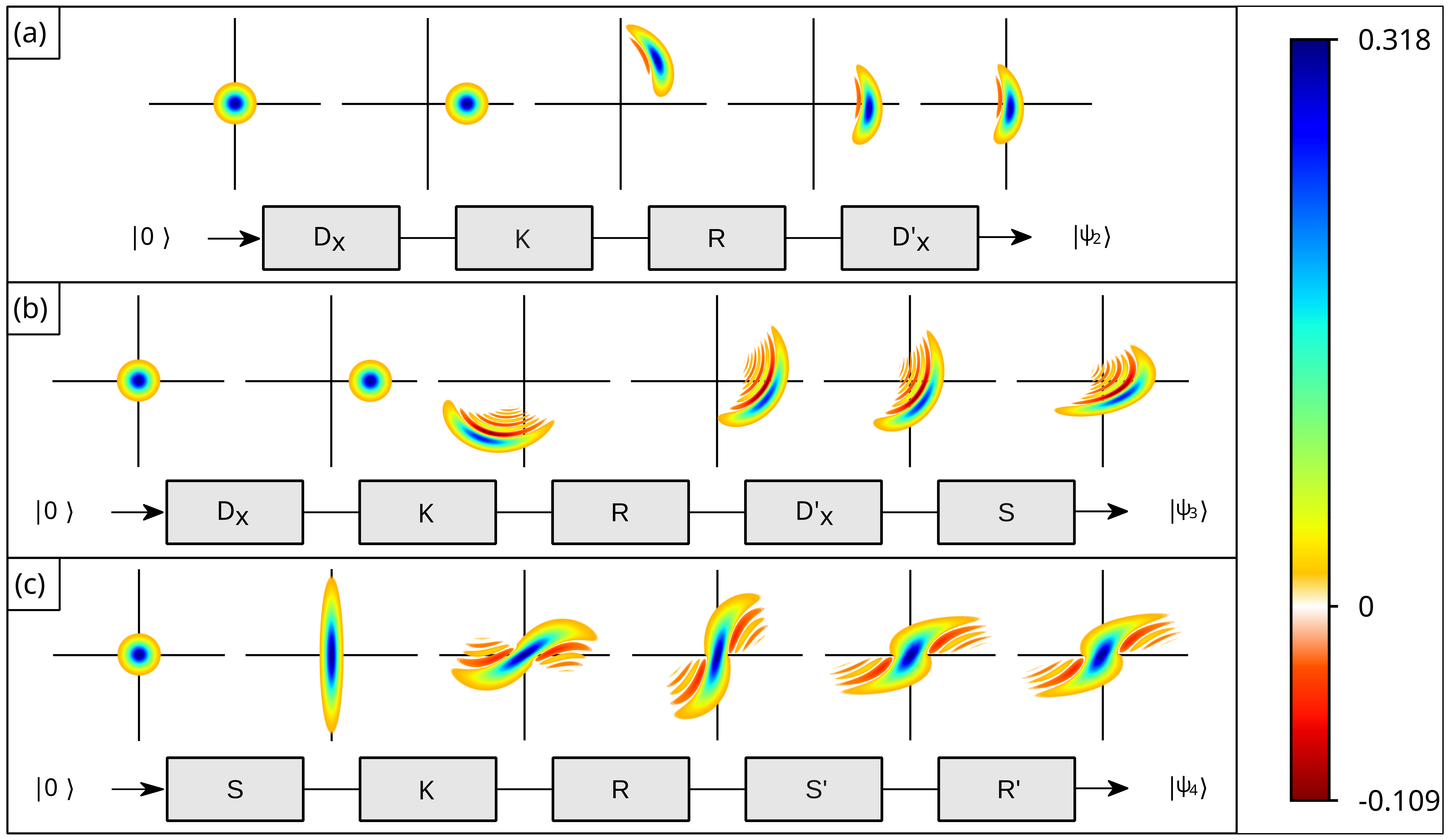}}
			\caption{\label{fig:Scheme} A schematic depiction of using the Kerr operation together with Gaussian gates for preparation of quantum states with: (a) linear squeezing, (b) nonlinear cubic squeezing, (c) nonlinear quartic squeezing. For better illustration the gates are supplemented by schematic depictions of the Wigner functions of the states along  different steps of the procedures, starting as the Wigner functions of vacuum states on the left, and ending as the Wigner functions of the respective squeezed states on the right. The boxes represent the unitary operations displacement (D$_{\text{x}}$, D$'_{\text{x}}$), phase shift (R), Kerr nonlinearity (K), and squeezing (S, S$'$). }
\end{figure}
Universal processing of continuous variable (CV) systems is defined as the ability to implement an arbitrary quantum operation \cite{Lloyd1999Feb}. This can be achieved by having an access to the class of the Gaussian operations that linearly transform the quadrature operators of the system \cite{Weedbrook2012May}, together with at least a single non-Gaussian operation \cite{Lloyd1999Feb, Mari2012Dec}. The non-Gaussian nature can be imparted by suitable projective measurements \cite{Ourjoumtsev2006Apr, Zavatta2004Oct}, but scalable applications ultimately demand a deterministic implementation. The unitary operation with the lowest order of a nonlinearity sufficient for the universal processing is the cubic phase gate \cite{Gottesman2001Jun}. In some systems the gate can be implemented by a direct dynamical control of the system's parameters \cite{Hillmann2020Oct, Cuevas2017Sep, Park2018May}, but it can be universally realized through a measurement induced scheme \cite{Marek2011Nov, Guo2015Feb, Miyata2016Feb, Marek2018Feb} with help of suitably prepared ancillary states.

The key property of the required ancillary states is the reduction of noise in the nonlinear quadrature corresponding to the cubic operation. The quantum states exhibiting such nonlinear squeezing are nonclassical and need to be prepared by specifically tailored techniques. For CV traveling light, this can be accomplished by preparing specific superpositions of photon number states by means of suitable projections by single photon detectors \cite{Konno2021Feb, Yukawa2013Nov}. Other physical systems, such as optomechanical systems \cite{Moore2019Nov}, microwave resonators \cite{Hillmann2020Oct}, or trapped ion systems \cite{Pedernales2015Oct,Park2018May}, can take an advantage of the ability to  dynamically control the coupling between the CV system and an ancillary mode. It is also possible to take advantage of the high-order nonlinearity that already exists in the physical system, such as suitably transforming quantum states produced by three-photon downconversion \cite{PRXQuantum.2.010327}.
Another prominent kind of nonlinearity, intensively pursued in a broad range of physical platforms, is the Kerr nonlinearity that nonlinearly affects phase of the system based on its energy. It is a non-Gaussian operation with broad applications in quantum logic \cite{Turchette1995Dec, Semiao2005Dec, Matsuda2007Oct, Azuma2007Dec}, quantum teleportation \cite{Vitali2000Jul, Jian2009Sep, Dong2013Apr}, or quantum non-demolition measurements \cite{Konig2002Oct, Xiao2008Dec}. The Kerr operation was already considered for a preparation of highly nonclassical superposed coherent states \cite{Jeong2004Dec} and, together with Gaussian operations, they can be employed for an incremental realization of the nonlinear operations of the third order \cite{Lloyd1999Feb}. It is therefore no surprise that it is being intensively studied across many scenarios, such as in electromagnetically induced transparency \cite{Schmidt1996Dec, xn--Imamolu-4s3c1997Aug, Werner1999Dec, Dey2007Jul}, Bose-Einstein condensates \cite{Hau1999}, cold atoms \cite{Kang2003Aug}, Josephson junctions \cite{Castellanos-Beltran2007Aug, Mallet2009Nov, Bergeal2010May}, and even light in resonators \cite{Niu2005Dec,Lu2013Oct}.

In this paper we show how can the Kerr nonlinearity be straightforwardly used for the generation of quantum states with the cubic and quartic nonlinear squeezing. Such states are the required resource for deterministic implementation of the quadrature phase gates and quantum nonlinear measurements. Even though their nonlinearity is unrelated to the Kerr operation, a single Kerr gate accompanied by the Gaussian processing is sufficient for deterministic preparation of the states. The state preparation procedure is fairly resistant to random fluctuations of the parameters which is a necessary requirement for the future experimental tests.

\section{Linear and nonlinear squeezing}
Squeezing is a process that reduces quantum fluctuations of a quantum operator. The process is a cornerstone for modern quantum technologies, it can be directly used for suppression of noise in quantum metrology \cite{Giovannetti2006Jan, Giovannetti2011Apr, Kwon2019Feb}, for generation of quantum entanglement \cite{Schumacher1995Apr,Kraus2008Oct, Bennett1996Apr, Fiurasek2002Sep}, or, in conjunction with non-Gaussian detectors, for preparation of highly nonclassical quantum states of light \cite{Ourjoumtsev2006Apr}. In general, squeezing is a nonclassical process that asymptotically transforms any quantum state into an eigenstate of the respective operator. The distinction between a linear and a nonlinear squeezing then falls down to the nature of the quantum operator. By the term linear squeezing we will denote the `traditional' Gaussian squeezing of quadrature operators $\hat{x}$ or $\hat{p}$ of a harmonic oscillator, or their linear combination. A straightforward application of squeezing operation  $\hat{S}(r) = \exp[ir(\hat{x}\hat{p} + \hat{p}\hat{x})/2]$  transforms the pair of these operators into $\hat{x}\rightarrow \hat{x}e^{r}$ and $\hat{p}\rightarrow \hat{p}e^{-r}$, where $r>0$ is a real parameter. The nonlinear squeezing then similarly transforms operators that are nonlinear functions of the quadrature operators \cite{Konno2021Feb}. The most straightforward example of the nonlinear squeezed states are the Fock states, $|n\rangle$, which are perfectly squeezed in the photon number operator $\hat{n} = (\hat{x}^2 + \hat{p}^2 - 1)/2$. However, in the following we will be more interested in the squeezing of quantum states which can be used as resources for the nonlinear quadrature phase gates that deterministically apply operation with the Hamiltonian $\hat{H}_n = \hat{p}^n$ \cite{Marek2018Feb}. Such states are squeezed in the operator $\hat{O}_n = \hat{x} - \hat{p}^{n-1}$ and the amount of squeezing in an arbitrary state $\hat{\rho}$ can be represented by ratio
\begin{equation}\label{xi_def}
    \xi_n = \frac{ \Tr[\hat{\rho} \hat{O}^2_n] - \Tr[\hat{\rho} \hat{O}_n]^2 }
    {\min_{\hat{\rho}_G}(\Tr[\hat{\rho}_G \hat{O}^2_n]-\Tr[\hat{\rho}_G \hat{O}_n]^2)},
\end{equation}
where the minimization in the denominator is taken over the set of all Gaussian states. When $\xi_n<1$ we can say that quantum state $\hat{\rho}$ has genuine nonlinear squeezing of the $n$-th order. It should be noted that quantum states squeezed according to (\ref{xi_def}) actually show the nonlinear squeezing of the respective order for a whole class of operators $\hat{O}_n = \hat{x} - \lambda \hat{p}^{n-1}$ with $\lambda>0$. This is because this parameter can be straightforwardly adjusted by the Gaussian squeezing operation which also affects the denominator of  (\ref{xi_def}) and the whole ratio then remains unchanged. The nonlinear squeezing is therefore a genuine non-Gaussian property of the quantum state.

The Kerr operation is unitary evolution operation determined by Hamiltonian $\hat{H} = \hat{n}^2$. However, in the following we shall employ an alternate form, $\hat{H}_K = \hat{n}^2 + 2 \hat{n} + 1 = ( \hat{x}^2 + \hat{p}^2 )^2 $, that has the advantage of being composed only of the quadrature operators of a single specific power, while differing from the original Kerr operation only by a phase shift that can be well controlled in quantum optics as well as on other experimental platforms. In terms of the quadrature operators, the Kerr operation is a nonlinear operation of the fourth order. A sequence of Kerr operations accompanied by suitable Gaussian operations can be therefore, in principle, used to create operations of any order \cite{Lloyd1999Feb}. In the following we will be interested in a more narrow task, namely the generation of quantum states with nonlinear squeezing. However, in turn, we will consider the limit of only a single Kerr operation accompanied by the Gaussian processing.

\section{Linear squeezing by Kerr operation}
\begin{figure}[h!]
		\centerline{\includegraphics[width=12cm]{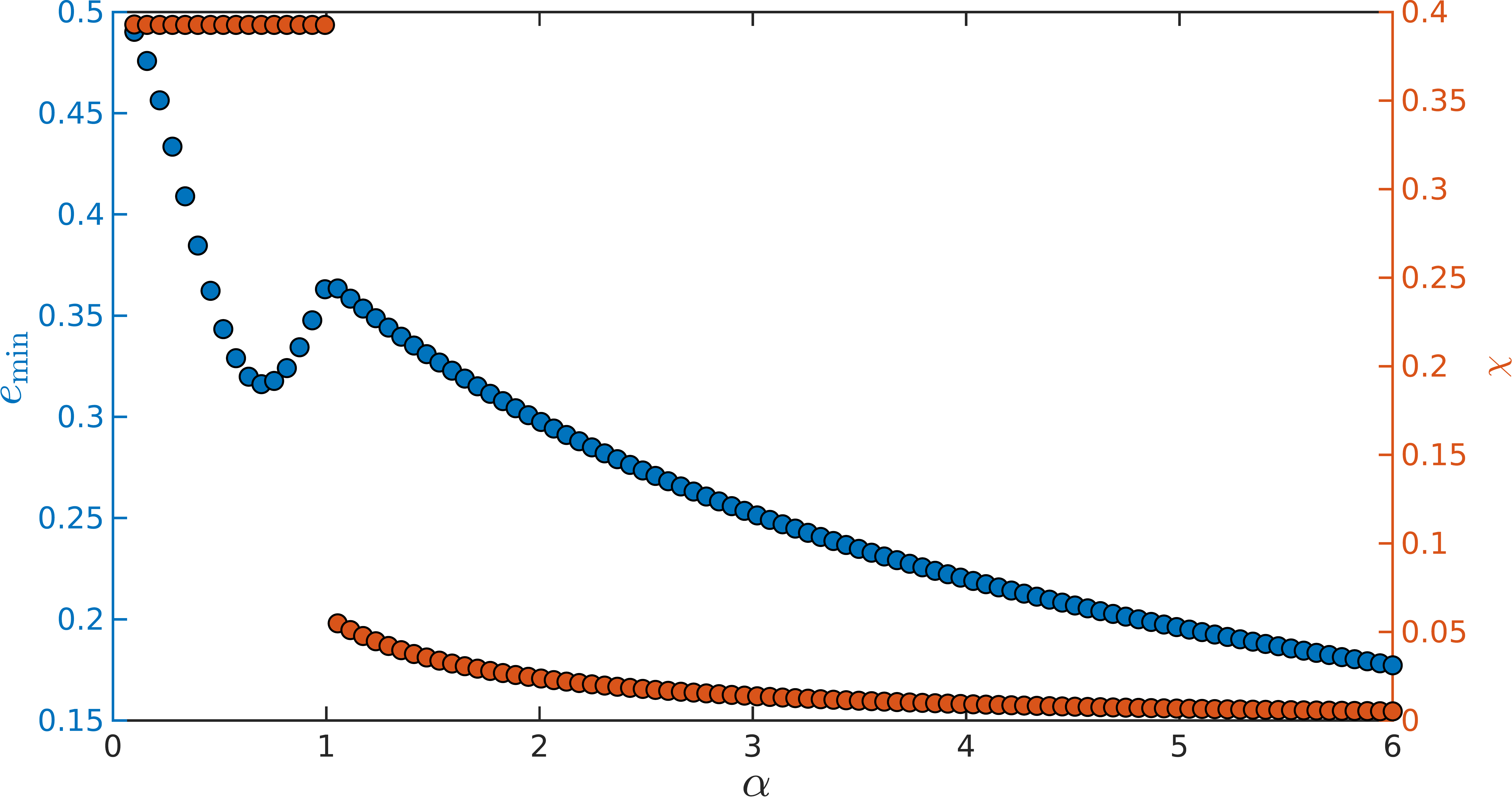}}
			\caption{\label{fig:Squeezing}
    An amount of the squeezing represented by the minimal eigenvalue of the variance matrix belonging to state (\ref{psi_x2}), (left y-axis, blue), and the value of the respective Kerr nonlinearity $\chi$ required for this result (right y-axis, red) relative to the displacement of the initial state $\alpha$.}
	\end{figure}

Let us start by looking at the generation of linear squeezing by using the Kerr operation. In the past this technique was employed for the generation of squeezed states \cite{Krivitsky2008Oct} and abandoned in favor of the second order parametric down-conversion. Here we consider it for instructive purposes. Since we are actively trying to generate a linearly squeezed quantum state, we will limit the available Gaussian operations only to those that cannot be used to implement squeezing directly - displacement and phase shift. Of these two, phase shift commutes with the Kerr operation and does not change the eigenvalues, so it can be ignored. The achievable squeezing is therefore fully determined by the second moments of quantum state
\begin{equation}\label{psi_x2}
		\ket{\psi_2} = \hat{K}(\chi)  \hat{D}_x(\alpha) \ket{0},
\end{equation}
where $\ket{0}$ is the vacuum state of the quantum harmonic oscillator, $\hat{D}_x(\alpha) = \exp(-i \alpha \hat{p})$ with $\alpha>0$ denotes the unitary displacement operation, and $\hat{K}(\chi) = \exp[-i\chi (\hat{x}^2 + \hat{p}^2)^2]$ denotes the Kerr operation. In any single mode quantum state, the amount of the squeezing can be evaluated from the least eigenvalue of the variance matrix of symmetrically ordered quadrature operators, which corresponds to the minimal variance that can be achieved by a suitable rotation in the phase space. The least eigenvalue $e_{\mathrm{min}}$ of the variance matrix belonging to state (\ref{psi_x2}), is fully given by the two real parameters $\chi$ and $\alpha$. Since there are only two parameters, we can numerically find the optimal $\chi$ for each $\alpha$. In this sense, the displacement parameter $\alpha$ supplies the quantum state with power, while the Kerr operation preserves the power and shapes it into the desired form. The results, which were obtained by numerical analysis in the truncated Fock space (see the supplement , Section 1. for more details), can be seen in Fig.~\ref{fig:Squeezing}.
There are several interesting observations that we can make. First, there are two regions of behavior distinguished by the energy of the initial state. In the first one, when $\alpha <1$, the required Kerr coefficient has a constant value and the maximal value of squeezing is achieved for a specific finite alpha. In the second region, corresponding to the initial states with larger energy, the achievable variance is monotonously decreasing. This trend suggests that in the limit of infinite energy the variance asymptotically approaches zero, even though it is impossible to check this only by the numerical tools. Also in this region of high energies, the required Kerr coefficient $\chi$ smoothly approaches value of zero. The difference can be understood by looking at the photon number distribution of the initial coherent state. For small values of $\alpha$, the dominant term of the initial quantum state can be, up to renormalization, expressed as superposition $|0\rangle + \alpha |1\rangle + \frac{\alpha^2}{\sqrt{2}}|2\rangle$. For this state the optimal Kerr operation introduces a sign flip between the $|1\rangle$ and the $|2\rangle$ terms, which is enough for the manifestation of squeezing. This changes when the $\alpha$ becomes larger and the higher photon number terms become relevant.

\section{Nonlinear squeezing by Kerr operation}
Let us now move towards generating states with the nonlinear squeezing. At first we consider states with cubic nonlinearity \cite{Konno2021Feb, Yanagimoto2020Jun, Hillmann2020Oct, Cuevas2017Sep}. Such states can be, in the idealized scenario, generated by applying a unitary cubic nonlinear operation given by Hamiltonian $\hat{H} = \hat{p}^3$ onto a Gaussian squeezed state. Such states always show the nonlinear squeezing as defined by the squeezing parameter $\xi_3 <1$ (\ref{xi_def}). Similarly to the case of a linear squeezing, the states with the cubic squeezing can be generated by applying the Kerr operation to an initial coherent state, as per (\ref{psi_x2}). However, in this case, the evaluation of the squeezing variance cannot be done by the straightforward computation of the variance matrix eigenvalues, because the nonlinear variance depends on higher than second moments. We therefore need to take into account the Gaussian operations, such as displacement, squeezing, and phase shift, which can be used to adjust the state after the Kerr operation. With such Gaussian processing, the final quantum state used for evaluation of the nonlinear variance can be read as
\begin{equation}\label{psi_x3}
    |\psi_3;\alpha,\chi,\phi,\beta,r \rangle = \hat{S}(r)\hat{D}(\beta)\hat{R}(\phi)\hat{K}(\chi)\hat{D}(\alpha)|0\rangle,
\end{equation}
where $\hat{S}(r) = \exp[ir(\hat{x}\hat{p} + \hat{p}\hat{x})/2]$ represents squeezing, $\hat{D}(\alpha) = \exp(-i \alpha \hat{p})$ represents displacement in $\hat{x}$ quadrature , $\hat{K}(\chi) = \exp[-i \chi (\hat{x}^2 + \hat{p}^2 )^2$] represents Kerr operation, and $R(\phi) = \exp[-i \phi (\hat{x}^2 + \hat{p}^2)/2]$ represents phase shift. The nonlinear squeezing parameter, $\xi_3(\alpha,\chi,\phi,\beta, r)$, is then a function of these five parameters which can be numerically optimized in order to obtain the largest nonlinear squeezing.

\begin{figure}[h!]
		\centerline{\includegraphics[width=12cm]{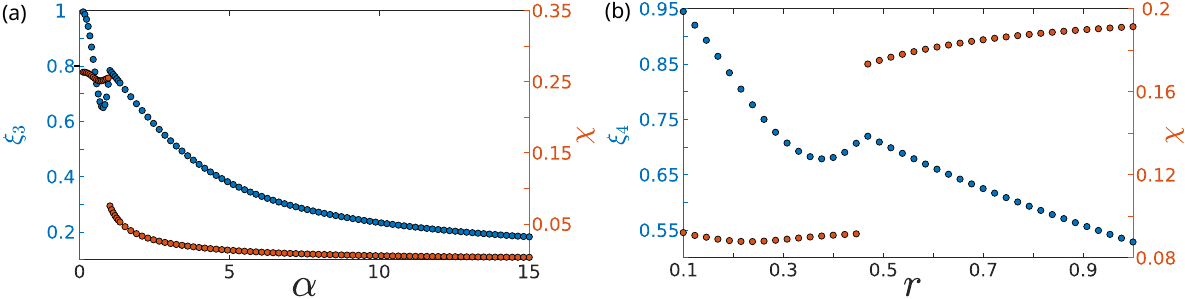}}
			\caption{\label{fig:Nonlinear}
			(a) The minimized cubic squeezing parameter $\xi_3$ (left y-axis, blue) and the $\chi$ coefficient of the Kerr nonlinearity required (right y-axis, red) relative to the displacement $\alpha$ of the initial state. (b) The minimized quartic squeezing parameter $\xi_4$ (left y-axis, blue) and the $\chi$ coefficient of the Kerr nonlinearity required (right y-axis, red) relative to the squeezing parameter $r$ of the initial state.}
	\end{figure}
We have performed such numerical optimization (for more details please see the supplement, section 2.) and the results are plotted in Fig.~\ref{fig:Nonlinear}(a). We can see that, on a qualitative level, the behavior is quite similar to generating the linear squeezing that's shown in Fig.~\ref{fig:Squeezing}.There are also two distinct regions of a behavior. The first is for the smaller displacements, $\alpha<1$, in which the required Kerr nonlinear coefficient is relatively large and mostly stable and the nonlinear squeezing attains a local minimum for a specific displacement amplitude. It is again the area in which the properties of the produced state are dominantly influenced by the coefficients of the first three Fock terms, $|0\rangle$, $|1\rangle$ and $|2\rangle$. It should be noted, though, that the nonlinear squeezing cannot be fully explained by these terms as the coherent state truncated to this dimension does not reach the values of Fig.~\ref{fig:Nonlinear}(a). The second region with $\alpha>1$ shows monotonous reduction of the nonlinear squeezing coefficient together with smooth diminishing of the required Kerr coefficient. As in the case of the linear squeezing, the cubic squeezing parameter $\xi_3$ seems to asymptotically approach zero, but the confirmation of this behavior goes beyond the range of our simulation parameters.

In a similar fashion we can evaluate generation of quantum states with the quartic squeezing, defined by having $\xi_4<1$ (\ref{xi_def}). Such states can be also prepared by a single Kerr operation, but it is no longer sufficient to start from a coherent state. The reason for this lies in the underlying symmetry of the state. Quantum states with the ideal cubic squeezing are symmetrical with respect to the change $\hat{p} \rightarrow -\hat{p}$, which geometrically represents symmetry with respect to a single axis in the phase space. In contrast, the states with ideal quartic squeezing have symmetry with respect to simultaneous exchange $\hat{x}\rightarrow -\hat{x}$ and $\hat{p}\rightarrow -\hat{p}$, which geometrically represents symmetry with respect to the point of the origin. The coherent states displaced in the $\hat{x}$ quadrature satisfy the first kind of a symmetry, which is the reason why they can be used for generation of the cubic squeezed states. Fortunately, the second kind of symmetry is satisfied by the Gaussian squeezed vacuum states. The approximation of a quantum state with the quartic squeezing can then be obtained as
\begin{equation}\label{psi_x4}
    |\psi_4;r,\chi,\phi_1,w,\phi_2\rangle = \hat{R}(\phi_2)\hat{S}(w)\hat{R}(\phi_1)\hat{K}(\chi)\hat{S}(r)|0\rangle.
\end{equation}
Similarly as in the previous analysis, the main nonlinear properties of the state are represented by the squeezing parameter $r$ and the Kerr parameter $\chi$. The other three parameters, phases $\phi_1$ and $\phi_2$, and second squeezing parameter $w$, represent the Gaussian processing with the purpose of adjusting the geometry of the quantum state so that the inherent nonlinear variance can be comfortably evaluated. As in the previous scenarios we consider the parameter $r$ to be the primary parameter determining the energy of the initial state and we can numerically optimize the other four parameters in order to find the highest quartic squeezing and the respective required Kerr parameter $\chi$. The results are plotted in Fig.~\ref{fig:Nonlinear}(b) and we can see that the behavior is qualitatively the same as in the previous scenarios. The different behavior for $r\lesssim 0.45$ is now dominantly influenced by the coefficients of the Fock states $|0\rangle$, $|2\rangle$, and $|4\rangle$, but similarly to the case of the qubic squeezing, the full state cannot be neglected as a truncated squeezed state cannot completely explain the values of the quartic squeezing in Fig.~\ref{fig:Nonlinear}(b).

The main difference between Fig.~\ref{fig:Nonlinear}(b) and Fig.~\ref{fig:Nonlinear}(a) is that in the case of the quartic squeezing, the required Kerr nonlinearity does not approach the value of zero, but it asymptotically increases towards $\chi \approx 0.2$. This is the consequence of the differences in photon number distributions of the two kinds of states. For coherent states, the weights of lower Fock states vanish as the energy increases, the Kerr operation affects only larger Fock states and can be accordingly smaller. For squeezed states the photon number distribution has always maximum for $\ket{0}$, the lower Fock states always remain relevant, and the Kerr coefficient therefore cannot be arbitrarily small.

\section{Error analysis}
We have shown that both the cubic and the quartic squeezed states can be generated by applying the Kerr operation to either coherent or squeezed states of the quantum harmonic oscillator. However, this process requires a precise control of all the parameters in the system. The initial quantum state needs to be prepared with the specific parameter, the Kerr operation needs to be applied with the specific interaction strength, and then the state needs to be post-processed in exactly the right way. In practical experiments, the ability to prepare the specific quantum states and to realize the specific quantum operations is not absolute. To verify the robustness of the state preparation protocol we have analyzed the proposed state preparation methods with respect to the random parameter fluctuations.
	\begin{figure}[h!]
		\centerline{\includegraphics[width=12cm]{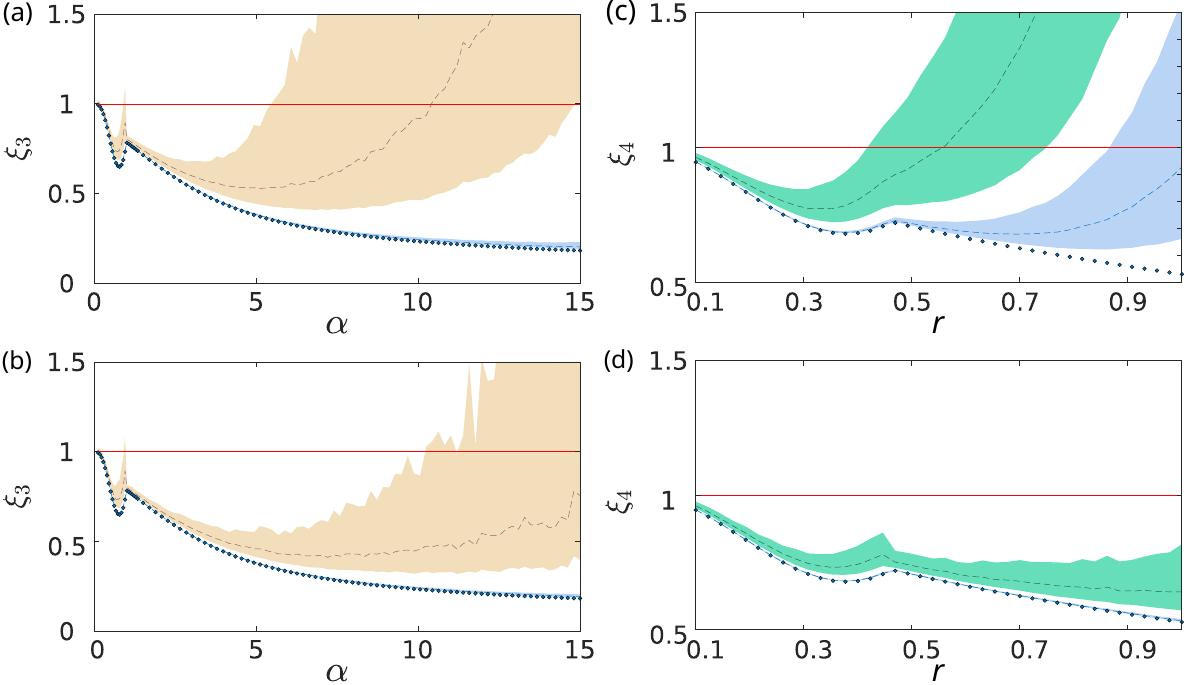}}
			\caption{\label{fig:Error} (a) Cubic nonlinear squeezing $\xi_3$ (\ref{xi_def}) for quantum state (\ref{psi_x3}) with parameters $\alpha$, $\chi$, $\phi$, $\beta$, and $r$  fluctuating with Gaussian distribution with the respective mean values $\mu_1,\cdots,\mu_5$ and with the standard deviations  $\sigma_1,\cdots,\sigma_5$. (b) Same as (a), but the initial displacements $\alpha$ are set to the optimal theoretical values. The red lines mark $\xi_3 =1$ and the areas below corresponds to quantum states with cubic nonlinear squeezing. The blue dots are representing the ideal scenario without the fluctuations. The dashed lines are showing the mean value of $\xi_3$ in the simulated sample, while the color filled areas mark the confidence interval between the upper and the lower standard deviation (see the supplement, Section 3. for more details). The red and the blue areas then mark the simulations with $\sigma_j = 0.05 \mu_j$ and $\sigma_j = 0.01 \mu_j$ for all $j$, respectively. (c) Quartic nonlinear squeezing $\xi_4$ (\ref{xi_def}) for a quantum state (\ref{psi_x4}) with parameters $r$, $\chi$, $\phi_1$, $w$, and $\phi_2$ fluctuating with Gaussian distribution with the respective mean values $\mu_1,\cdots,\mu_5$ and with the standard deviations  $\sigma_1,\cdots,\sigma_5$. (d) Same as (c), but the Kerr interaction strengths $\chi$ are set to the optimal theoretical values. The red lines mark $\xi_4 =1$ and the areas below correspond to states with the quartic nonlinear squeezing. The blue dots represent the ideal scenario without the fluctuations. The dashed lines are showing the mean value $\xi_4$ in the simulated sample, while the color filled areas mark span between the upper and the lower standard deviation (see the supplement, Section 3. for more details). The green and the blue areas then mark the simulations with $\sigma_j = 0.05 \mu_j$ and $\sigma_j = 0.01 \mu_j$ for all $j$, respectively.}
	\end{figure}

Since the linear squeezing can be straightforwardly implemented without employing the Kerr nonlinearity, we focused our attention on preparation of states with the nonlinear squeezing. Let us start with the cubic nonlinear squeezing. According to (\ref{psi_x3}), there are five parameters that determine the properties of the state: the initial displacement, and the strength of Kerr nonlinearity, rotation, displacement and squeezing that have values found by the optimization algorithm. To analyze the errors arising from the imprecise control we have numerically simulated 10 000 runs of the experiment, each one with the five-tuple of parameters randomly generated with respective Gaussian distributions with the mean values $\mu_1,\cdots,\mu_5$ equal to the ideal theoretical values, and the standard deviations $\sigma_1,\cdots,\sigma_5$, corresponding to specific fractions of the respective mean values, either $\sigma_j = 0.01 \mu_j$ or $\sigma_j = 0.05\mu_j$, for all $j$. We have evaluated the nonlinear cubic squeezing of the quantum states produced in the individual simulated runs and we have aggregated these results to obtain the statistics shown in Fig.~\ref{fig:Error}(a). In the figure, the dashed lines, red for $\sigma_j = 0.01 \mu_j$ and blue for $\sigma_j = 0.05\mu_j$, show the mean values of the generated nonlinear squeezing, while the colored area marks the interval with a width of one standard deviation of the produced $\xi_3$ (see the supplement, Section 3. for more details). We can see that while the $1\%$ errors with $\sigma_j = 0.01\mu_j$ do not significantly hamper the generation of the cubic squeezing, the $5\%$ errors can already prevent cubic squeezing appearing beyond the initial dip. It should be noted that deterioration of the average value of the cubic nonlinear squeezing is mostly caused by a small percentage of the outliers - in the case of the $5\%$ errors, $77.4\%$ of the simulated runs have shown the nonlinear squeezing below the mean value. The influence of the significant outliers then could be, in principle, conditionally reduced by the methods of quantum state purification, see for example \cite{Glockl2006Aug}. We have also checked the relative importance of the five parameters in (\ref{psi_x3}) by performing a series of the simulated runs, each with one  parameter affixed and others selected randomly. In the case of the cubic nonlinear squeezing, the initial squeezing had the strongest influence on the final result. With the initial fixed displacement and the initial pure coherent state, the resulting nonlinear squeezing is shown in Fig.\ref{fig:Error}(b) and there we can see that even the $5\%$ errors generate the states that show, on average, the nonlinear squeezing across the full set of the chosen values.

We have proceeded similarly in the second case dealing with preparation of the quartic nonlinear squeezed states with $\xi_4<1$ (\ref{xi_def}). Again, according to (\ref{psi_x4}), there are five parameters determining the nonlinear squeezing of the state: the initial squeezing parameter $r$, and the parameters for the Kerr nonlinearity, rotation in phase space, another squeezing and another rotation that were optimized in order to achieve the minimum in $\xi_4$ of the output state. The errors caused by the fluctuations in these parameters were evaluated again by numerically simulating 10000 runs of the experiment with the Gaussian random fluctuations of all the parameters and the results are shown in Fig.\ref{fig:Error}(c). We can see that the quartic squeezing is significantly more vulnerable to the imperfections as even for the errors with $\sigma_j= 0.01\mu_j$, the quartic squeezing does not surpass $\xi_4\approx 0.7$, that can be achieved only by considering superposition of the Fock states $|0\rangle$ and $|2\rangle$. While $61\%$ of all the data points lie under the mean value for $\sigma_j = 0.05\mu_j$, and $67\%$ of all data points lie under the mean value for $\sigma_j = 0.01\mu_j$, the outliers show the values large enough to practically prevent the generation of the quartic nonlinear squeezing in this fashion. The analysis of a relative importance of the individual parameters revealed that the determining parameter is the strength of the Kerr nonlinearity $\chi$. When it is fixed, which is the case for the scenario shown in  Fig.\ref{fig:Error}(d), the scenario with $1\%$ errors closely matches the ideal setting and even $5\%$ errors lead to consistently decreasing values of $\xi_4$.

\section{Conclusion}
In principle, quantum states exhibiting nonlinear squeezing can be generated with help of an arbitrary high-order nonlinearity through the geometric phase effect \cite{Lloyd1999Feb}. However, such approach is incremental and requires a large number of individual nonlinear operations to achieve the desired result. We have shown that states exhibiting the nonlinear squeezing of the third and the fourth order can be generated with the help of only a single Kerr gate with a constant interaction strength and a set of suitably chosen Gaussian operations. In both cases, the key step is applying the Kerr gate onto a Gaussian quantum state - a coherent state for the cubic operation and a squeezed state for the quartic operation, the particular choice being determined by the symmetry of the required nonlinear squeezing. In both cases, some nonlinear effect can be already obtained from considering the first three nonzero terms in the Fock state representation of the quantum states, but this can explain only part of the nonlinear effect and taking advantage of the full Hilbert space is therefore always beneficial.

Successfully preparing the desired quantum states requires a precise alignment of all parameters of both the Gaussian and the non-Gaussian constituent operations. To test the experimental viability of the proposed operations we have to numerically analyzed their performance under the fluctuations of these parameters. Gaussian fluctuations in all parameters with the standard deviations on the order of $5\%$ of the means for the cubic states and $1\%$ of the means for the quartic states can be roughly tolerated. We therefore expect that this technique could be experimentally tested on the platforms on which is the Kerr gate currently available \cite{Schmidt1996Dec, xn--Imamolu-4s3c1997Aug, Werner1999Dec, Dey2007Jul,Hau1999,Kang2003Aug,Castellanos-Beltran2007Aug, Mallet2009Nov, Bergeal2010May,Niu2005Dec,Lu2013Oct}.

\section*{Funding}
This work was supported by the Czech Science Foundation, grant No. GA18-21285S. We also acknowledge the European Union's Horizon 2020 research and innovation programme (CSA Twinning) under Grant Agreement No 951737 (NONGAUSS). \v{S}. Br\"{a}uer acknowledges internal support by Palack\'{y} University through the project IGA-PrF-2021-006.

\section*{Acknowledgements}
We acknowledge an use of cluster computing resources provided by the Department of Optics, Palack\'{y} University Olomouc. We thank to J. Provazn\'{i}k for maintaining the cluster and providing support.

\newpage

\section*{Supplement}
\setcounter{section}{0}

\section{Optimal Gaussian states for evaluating nonlinear squeezing.}
In chapter IV we are dealing with the nonlinear squeezing and the minimization of $\xi_{3,4}$. But in order to minimize $\xi_{3,4}$ we first need the optimal variance of the ideal Gaussian state in the denominator. The ideal Gaussian state for the cubic nonlinearity is the squeezed state, so we have analytically calculated the optimal squeezing parameter for a minimal variance, then we did it in the Heisenberg's picture, where $\hat{x}$ and $\hat{p}$ evolve as
	\begin{align}
		\hat{x} \rightarrow & g\hat{x},\\
		\hat{p} \rightarrow & \frac{1}{g} \hat{p},
	\end{align}
	then we substitute the new $\hat{x}$ and $\hat{p}$ into the variance
	\begin{align}
		\text{var} \left(\hat{O}_3\right) =& \langle \hat{O}_3 ^2 \rangle - \langle \hat{O}_3 \rangle ^2; \text{~} \hat{O}_3 = \hat{x} -\hat{p}^2 , \\
	\langle \hat{O}_3 \rangle =& g \bra{0} \hat{x} \ket{0} - \frac{1}{g^2} \bra{0} \hat{p}^2 \ket{0},\\
	\langle \hat{O}_3^2 \rangle =& g^2 \bra{0} \hat{x}^2 \ket{0} - \frac{1}{g} \bra{0}\hat{x}\hat{p}^2 + \hat{p}^2\hat{x} \ket{0} +\nonumber \\ +& \frac{1}{g^4} \bra{0} \hat{p}^4 \ket{0},\\
		\text{var}\left(\hat{O}_3 \right) =& \frac{g^2}{2} + \frac{1}{g^4}\frac{3}{4} - \frac{1}{g^4}\frac{1}{4} = \frac{g^2}{2} + \frac{1}{2 g^4}.
	\end{align}
The next step is to perform a partial derivative and to set it equal to the zero value for finding the extreme value.
	\begin{align}
		\frac{\partial \text{var}\left( \hat{O}_3 \right)}{\partial g} =& 0 ,\\
		g - \frac{2}{g^5} = 0 \rightarrow & g = \sqrt[6]{2}, \\
		\text{var}_{min} \left( \hat{O}_3 \right) =& 3 \times 2 ^{\frac{5}{3}}	.	
	\end{align}
	
We have applied the same procedure in the case of the quartic squeezing, where the quadratures evolve as
	\begin{align}
		\hat{x} \rightarrow & g \cos(\phi) \hat{x} + \frac{\sin(\phi)}{g}\hat{p},\\
		\hat{p} \rightarrow &\frac{\cos(\phi)}{g}\hat{p} - g \sin(\phi) \hat{x}.
	\end{align}
Subsequently, we substitute that into the variance
	\begin{align}
		\text{var}\left( \hat{O}_4 \right) =& \langle \hat{O}^2 _4 \rangle - \langle \hat{O}_4 \rangle ^2;\text{~} \hat{O}_4 = \hat{x} - \hat{p}^3 ,\\
		\text{var}\left( \hat{O}_4 \right) =& \frac{g^2 \cos^2 (\phi)}{2} + \frac{\sin^2 (\phi)}{2g^2} + \nonumber\\
	    +& \frac{3}{2} g^2 \sin^3(\phi)\cos(\phi) + \frac{3}{2} \sin(\phi) \cos^3 (\phi) -\nonumber\\
	    -& \frac{3}{2} \sin^3(\phi) \cos(\phi)  - \frac{3}{2} \frac{\cos^3(\phi) \sin(\phi)}{g^4} +\nonumber\\     +& \frac{15}{8} \frac{\cos^6(\phi)}{g^6} + \frac{45}{8}\frac{\sin^2(\phi) \cos^4(\phi)}{g^2} +\nonumber\\ +& \frac{45}{8} g^2 \sin^4(\phi) \cos^2(\phi) + \frac{15}{8} g^6 \sin^6 (\phi).
	\end{align}
The optimal parameters for the Gaussian state minimizing this formula can be numerically found to be
	\begin{align}
		g =& -0.637, \\
		\phi =&  -1.949,
	\end{align}
and the minimal variance of the Gaussian state is $\text{var}_{min}\left( \hat{O}_4 \right) = 0.971$.

\section{Optimization of the state preparation}
In the following we would like to describe the numerical optimization tools employed to achieve the results in the paper.
The optimized parameter is always the relative squeezing parameter $\xi_{3,4}$, but since the denominator is always fixed, as shown in Appendix A, it is sufficient to minimize the variance of the quantum operators $\hat{O}_3$ and $\hat{O}_4$ with $\hat{O}_n = \hat{x}-\hat{p}^{n-1}$. These variances, which we will denote $V_3(\alpha; \chi, \phi,\beta, r)$ and $V_4(r; \chi, \phi_1,\omega, \phi_2)$, respectively, are the functions of the input parameters of the preparation circuit.

The variances for each combination of parameters were calculated in two steps. In the first step we numerically applied the Kerr nonlinearity by expressing the input state in the Fock basis of the dimension $N$ and by multiplying it by  matrix form of the unitary operator for the Kerr operation to produce the approximate representation of the states $|\zeta_3\rangle = \hat{K}(\chi)\hat{D}(\alpha)|0\rangle$ and $|\zeta_4\rangle = \hat{K}(\chi)\hat{S}(r)|0\rangle$ for the preparation of the cubic and the quartic squeezing, respectively.
The application of the Kerr operator could be done perfectly because the operator is diagonal in the Fock basis. In our simulations we have used dimension $N = 300$ which was sufficient to faithfully represent the selected input states. We then used these quantum states to evaluate the moments of the quadrature operators.

For generation of the cubic squeezing, the relevant moments can be obtained by the Gaussian transformation of the quadrature operators. That leads to the polynomial formula for the variance
\begin{equation}
		V_3 = \bra{\zeta_3} (\hat{O}'_3)^2 \ket{\zeta_3} -  \bra{\zeta_3} \hat{O}'_3 \ket{\zeta_3}^2,
\end{equation}
where
\begin{align}
		\hat{x}' =& g  \left( \cos(\phi)x + \sin(\phi)p \right),\\
		\hat{p}' =& \frac{1}{g} \left( (-\sin(\phi)x +\cos(\phi)p) + \beta \right),\\
		\hat{O}'_3 =& \hat{x}' - \hat{p}'^3.
\end{align}
Similarly for the quartic squeezing we obtain
\begin{equation}
    V_4 = \bra{\zeta_4} (\hat{O}'_4)^2 \ket{\zeta_4} -  \bra{\zeta_4} \hat{O}'_4 \ket{\zeta_4}^2,
\end{equation}
where
\begin{align}
		\hat{x}_1 =& \omega \sin(\phi_2) \left( \sin(\phi_1)\hat{x} + \cos(\phi_1)\hat{p}\right), \\
		\hat{x}_2 =& \frac{1}{\omega} \cos(\phi_2) \left(- \cos(\phi_1)\hat{x} + \sin(\phi_1)\hat{p}\right), \\
		\hat{p}_1 =& \frac{1}{\omega} \sin(\phi_2) \left(- \cos(\phi_1)\hat{x} + \sin(\phi_1)\hat{p}\right), \\
		\hat{p}_2 =& \omega \cos(\phi_2) \left( \sin(\phi_1)\hat{x} + \cos(\phi_1)\hat{p}\right), \\
		\hat{x}" =& \hat{x}_1 + \hat{x}_2,\\
		\hat{p}" =& \hat{p}_1 - \hat{p}_2,\\		
		\hat{O}_4 =& \hat{x}" - {\hat{p}"}^3.
\end{align}

	\begin{figure}[h!]
		\centerline{\includegraphics[scale=1]{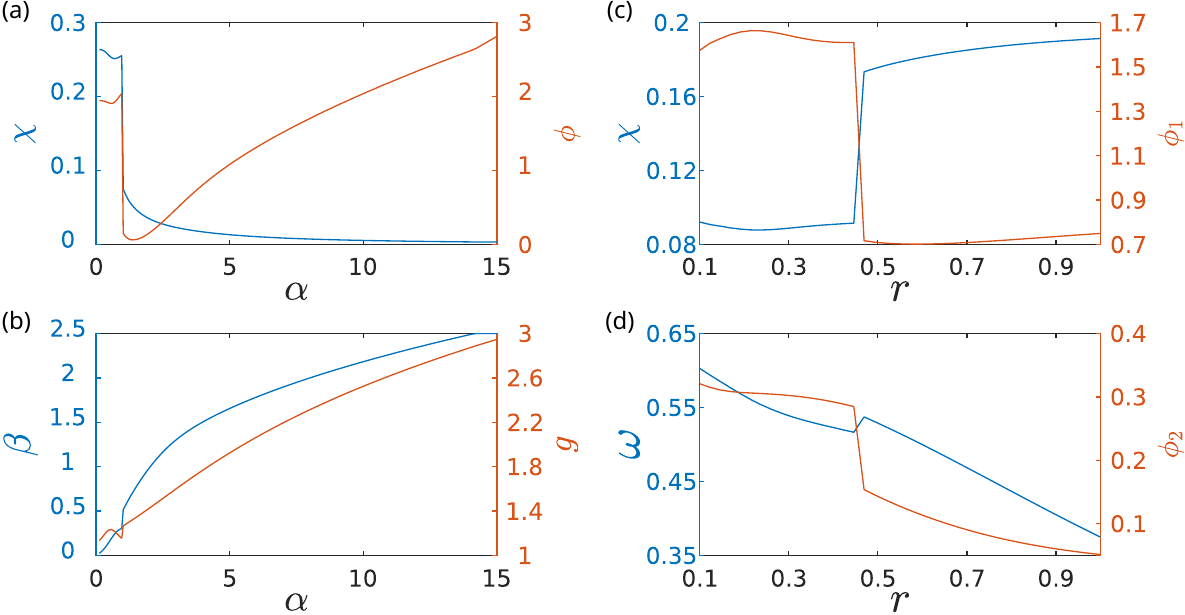}}
			\caption{\label{fig:Parameters} Left column - the optimal parameters of obtained by the numerical optimization for the given input displacement for the case of the Cubic nonlinear squeezing. (a) The parameters of Kerr nonlinearity $\chi$, and phase shift parameter $\phi$. (b) The displacement parameter $\beta$, and the squeezing parameter, $g$. Right column - the optimal parameters obtained by the numerical optimization for a given input displacement for the case of the Quartic nonlinear squeezing. (c) The parameters of Kerr nonlinearity $\chi$, and rotation $\phi_1$. (d) The phase shift parameter $\phi_2$, and the squeezing parameter $(\omega)$.}
	\end{figure}

The numerical optimization of the functions was performed in  Python, with help of the {\it{scipy.optimize.minimize}} library and the L-BFGS-B function. This is the quasi-Newtonian optimization method that allows to set the intervals of parameters in which the optimization will take place and thus reduce the computational time. This method uses the Broyden-Fletcher-Goldfarb-Shanno algorithm \cite{fletcher1987practical}. The optimization searches for a local minima and always starts from the pre-selected entry points. In our analysis, in which we searched for the minimal values for the different fixed values of $\alpha$ (for the cubic squeezing) and $r$ (for the quartic squeezing), we always chose one of the sets of the parameters as those that were optimal for the previously calculated value of $\alpha$ or $r$, and the other 299 sets were chosen randomly. Together there were 300 different starting sets of the parameters for an each instance of the numerical optimization.

The parameters for which the optimal nonlinear squeezing  was found are plotted for the cubic squeezing in Fig.~\ref{fig:Parameters}(a),(b) and for the quartic squeezing in Fig.~\ref{fig:Parameters}(c),(d).

\section{Statistical evaluation of errors}
In the following we would like to present the detailed explanation of the error analysis which was presented in Section V. When simulating the errors, we start from the optimal set of the parameters and then simulate the random deviations. This is done by running the Monte Carlo simulation, in which $N_{runs} = 10000$ runs of the quantum state preparation are simulated with parameters that are randomly chosen from a Gaussian distribution with the mean values $\mu$, that's corresponding to the parameter's optimal value and the standard deviations $\sigma = \gamma \mu$, which are considered to be a certain fraction of the mean value. This represents the inability to precisely control the quantum systems. Then the obtained nonlinear variances from an each run are statistically evaluated.

In each simulated run of the experiment, the obtained nonlinear variance can be expressed as $\xi_{3,4}(k)$, where $k = 1,\cdots,N_{runs}$ denotes the particular run. The fundamental information is provided by the statistical moments. The most important is the mean value, $\overline{\xi}_{3,4}  = \frac{1}{N_{runs}}\sum_k \xi_{3,4}(k)$, but important insight is also given by the upper and the lower standard deviations
\begin{align}\label{}
    \sigma_+^2 =& \frac{1}{N_+}\sum_{k} (\mathrm{max}[\xi_{3,4}(k) - \overline{\xi}_{3,4},0])^2 ,\\
    \sigma_-^2 =& \frac{1}{N_-}\sum_{k} (\mathrm{min}[\xi_{3,4}(k) - \overline{\xi}_{3,4},0])^2 ,
\end{align}
where $N_+$ and $N_-$ represent the number of runs in which the measured nonlinear variance $\xi_{3,4}(k)$ is larger or lower, respectively, than the mean variance $\overline{\xi}_{3,4}$.


\end{document}